\def\year{2019}\relax
\documentclass[letterpaper]{article} 
\usepackage{aaai19}  
\usepackage{times}  
\usepackage{helvet}  
\usepackage{courier}  
\usepackage{url}  
\usepackage{graphicx}  
\usepackage{amsmath}
\usepackage{amssymb}
\usepackage{algpseudocode}
\usepackage{algorithm}
\usepackage{caption}
\usepackage{subcaption}
\usepackage{multirow}
\usepackage{url}
\usepackage{color}

\newtheorem{mydef}{Definition}

\long\def\comment#1{}

\title{Unsupervised Meta-path Reduction on Heterogeneous Information Networks}

\author{Xiaokai Wei, Zhiwei Liu, Lichao Sun, Philip S. Yu\\
	Department of Computer Science, University of Illinois at Chicago\\ weixiaokai@gmail.com, \{zliu213, lsun29, psyu\}@uic.edu}
 
\begin{document}
%

\maketitle
\begin{abstract}  \small\baselineskip=9pt
	Heterogeneous Information Network (HIN) has attracted much attention due to its wide applicability in a variety of data mining tasks, especially for tasks with multi-typed objects. A potentially large number of meta-paths can be extracted from the heterogeneous networks, providing abundant semantic knowledge. However, though a variety of meta-paths can be defined, too many meta-paths are redundant. Reduction on the number of meta-paths can enhance the effectiveness since some redundant meta-paths provide interferential linkage to the task. Moreover, the reduced meta-paths can reflect the characteristic of the heterogeneous network. Previous endeavors try to reduce the number of meta-paths under guidance of supervision information. Nevertheless, supervised information is expensive and may not always be available. In this paper, we propose a novel algorithm, SPMR (\textbf{S}emantic \textbf{P}reserving \textbf{M}eta-path \textbf{R}eduction), to reduce a set of pre-defined meta-paths in an unsupervised setting. The proposed method is able to evaluate a set of meta-paths to maximally preserve the semantics of original meta-paths after reduction. Experimental results show that SPMR can select a succinct subset of meta-paths which can achieve comparable or even better performance with fewer meta-paths.
\end{abstract}

\section{Introduction}

\begin{figure*}[t!]
	\centering
	\begin{subfigure}[t]{0.25\textwidth}
		\centering
		\includegraphics[height=1in]{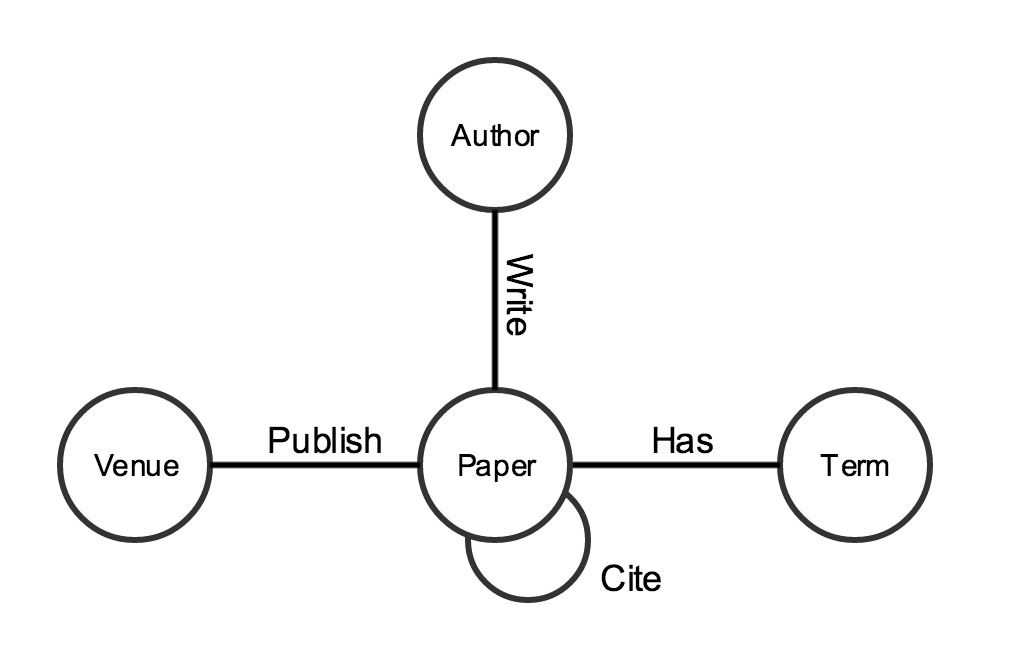}
		\caption{DBLP}
		\label{fig:dblp}
	\end{subfigure}%
	~
	\begin{subfigure}[t]{0.25\textwidth}
		\centering
		\includegraphics[height=1in]{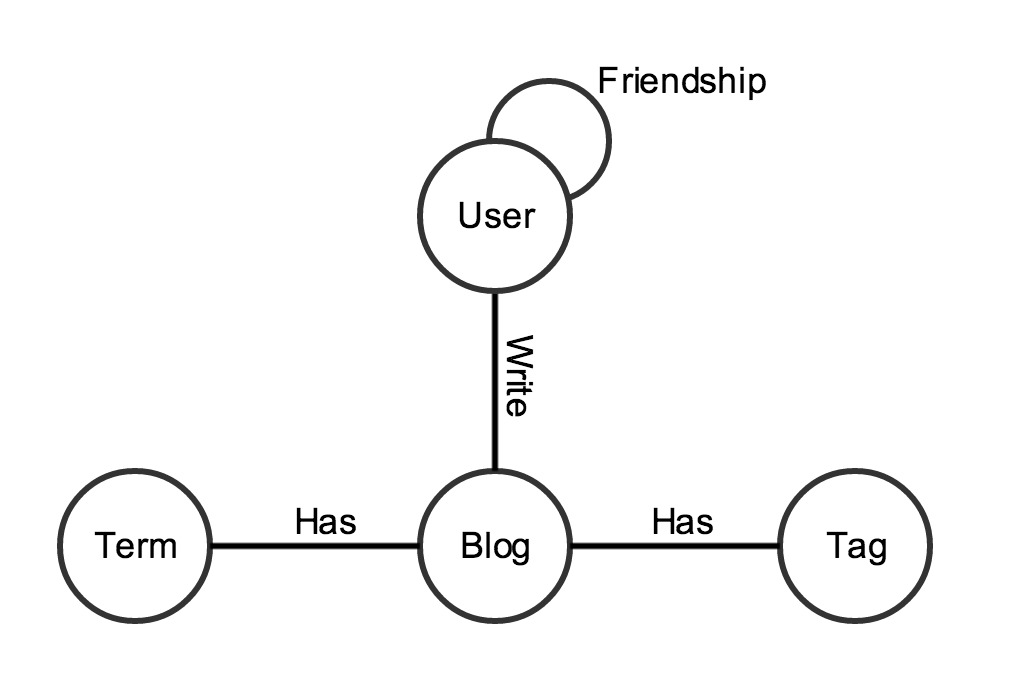}
		\caption{BlogCatalog}
		\label{fig:blog}
	\end{subfigure}
	~
	\begin{subfigure}[t]{0.25\textwidth}
		\centering
		\includegraphics[height=1.6in]{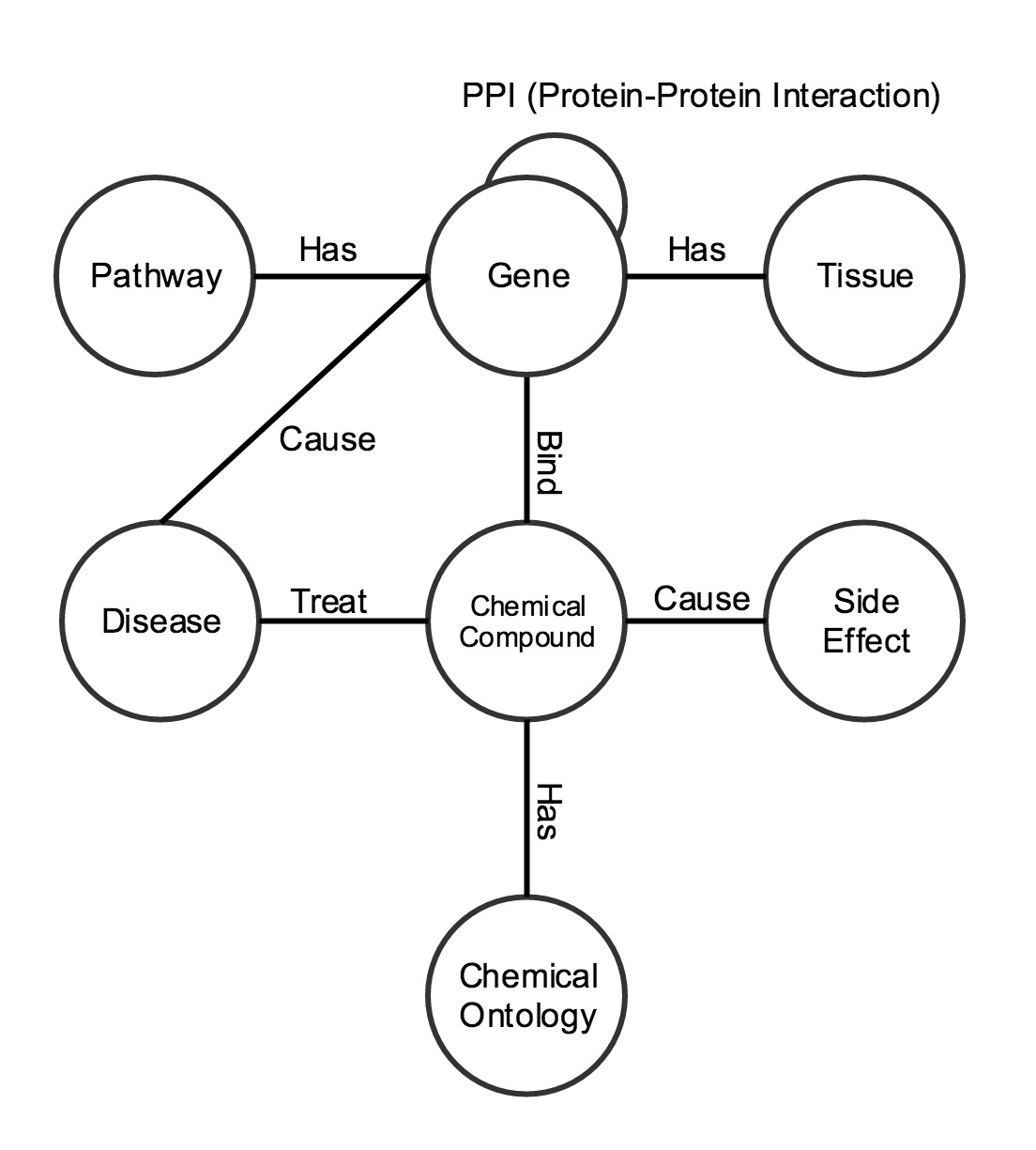}
		\caption{Bioinformatic}
		\label{fig:gene}
	\end{subfigure}
	\caption{Examples of heterogeneous information network}
	\label{fig:network}
\end{figure*}

Information networks, such as friendship networks and gene networks, have been widely studied in various data mining tasks, such as community detection \cite{Leskovec10} and collective classification \cite{Lu03}. Traditional information networks usually assume the nodes are of the same type, and such networks are usually referred to as {\it homogeneous networks}. However, in the era of big data, different types of real-world objects from various domains are often inter-connected. For example, in bibliographic network (e.g., DBLP\footnote{\url{http://dblp.uni-trier.de/}}), {\it author}, {\it paper}, {\it term} and {\it venue} constitute a multi-typed network in which different types of relationship (Figure \ref{fig:dblp}) exist among nodes (e.g., {\it author} 'writes' {\it paper}, {\it paper} 'contains' {\it term}). In social media network (e.g., Twitter/BlogCatalog), objects including {\it tweet/blog}, {\it user}, {\it hashtag/tag} and {\it term} also interact with each other (Figure \ref{fig:blog}). {\it Heterogeneous Information Network} (HIN) \cite{Sun12},\cite{sun2018mega} has been proposed to model such interacting multi-typed objects. Due to its versatility in modeling inter-connected objects, HIN has been employed in a wide variety of applications, including recommendation \cite{Yu14}, classification \cite{Kong12} \cite{Kong13a}, clustering \cite{Sun12} and information fusion \cite{Kong13} \cite{Zhang14}.

In order to utilize the rich information embedded in HIN, one popular way is to extract meta-paths \cite{Sun09} from the heterogeneous network. Meta-path is a sequence of relations which captures the correlation among object types.  Generally, there are a variety of meta-paths in a heterogeneous network. Applying all types of meta-paths at the same time may lead to low-efficiency problem. Moreover, some meta-paths may carry misleading information, known as social noise~\cite{liu2017event}, which can become an interference to the tasks. According to our experiment, a small subset of meta-paths can provide sufficient information. Hence, it is desirable to reduce the number of meta-paths so that the performance could be better and the characteristic of HIN can be revealed explicitly.

In a supervised scenario, the reducing process of the number of meta-paths is relatively easy, as supervision can be used as a guidance for weighting different meta-paths. The supervision information can be user (implicit) feedback for recommendation problem \cite{Yu14} or class label for classification problem \cite{Kong13}. The meta-paths having higher correlation with the supervision information can be retained and the meta-paths with little correlation can be discarded.

However, supervision information is not always available and is usually expensive to obtain. In this paper, we propose a novel approach to implement reduction on meta-paths under unsupervised setting, which is non-trivial due to the lack of guidance. This approach aims to reduce the number of a pre-defined set of meta-paths while still preserving the semantic information of the original network. Hence, the reduced subset of meta-paths can be viewed as a succinct summary of the original meta-paths. The performance of subsequent task could also be enhanced since the abandoned meta-paths may constitute the noise part of the network. Furthermore, the reduced set of meta-paths can provide human analysts better insights about the characteristic of the network.


The main contribution of this paper can be summarized as follows:
\begin{itemize}
	\item To our best knowledge, we are the first to formally study the problem of unsupervised meta-path reduction on heterogeneous information networks. We aim to select a succinct subset of meta-paths which can preserve most of the information of all the meta-paths.
	\item We propose a transition probability preserving approach to perform meta-path reduction, which utilizes correlation among different meta-paths. 
	\item We conduct experiments on two real-world datasets to show that our proposed methods can perform better in terms of the clustering accuracy and the reduced set can still preserve the semantic information.
\end{itemize}

The rest of the paper is organized as follows. We present preliminary concepts in section 2 and propose the approach in section 3. Optimization one the proposed model is introduced in section 4. In section 5, experimental results are shown to compare the proposed approach with using all the meta-paths. And we review some related work in section 5 before we conclude our work in section 5.

\begin{table*}
	\begin{center}
		\caption{Examples of meta-paths derived from two datasets}
		\begin{tabular}{| l | l | }
			\hline
			Datasets & Examples of meta-path \\ \hline
			\multirow{3}{*}{BlogCatalog} & Blog $\xrightarrow{has}$ Tag $\xrightarrow{has^{-1}}$ Blog \\
			& Blog $\xrightarrow{written\_by}$ User $\xrightarrow{written\_by^{-1}}$ Blog \\
			& Blog $\xrightarrow{written\_by}$ User $\xrightarrow{friend}$ User $\xrightarrow{written\_by^{-1}}$ Blog \\ \hline
			\multirow{3}{*}{DBLP} & Paper $\xrightarrow{has}$ Term $\xrightarrow{has^{-1}}$ Paper \\
			& Paper $\xrightarrow{written\_by}$ Author  $\xrightarrow{written\_by^{-1}}$ Paper \\
			& Paper $\xrightarrow{written\_by}$ Author $\xrightarrow{written\_by^{-1}}$ Paper $\xrightarrow{has}$ Term $\xrightarrow{has^{-1}}$ Paper  \\ \hline
			\multirow{6}{*}{Chemical Compound}
			&   Compound$\xrightarrow{bind}$ Gene $\xrightarrow{PPI}$ Gene $\xrightarrow{bind^{-1}}$Compound \\
			&   Compound$\xrightarrow{treat}$Disease$\xrightarrow{cause^{-1}}$Gene$\xrightarrow{bind^{-1}}$ Compound \\
			&   Compound $\xrightarrow{bind}$ Gene $\xrightarrow{has }$ Pathway $\xrightarrow{has^{-1}}$ Gene $\xrightarrow{bind^{-1}}$ Compound \\
			\hline
		\end{tabular}
		\label{table:metapath}
	\end{center}
\end{table*}

\section{Related Work}

In this section, we review some related work on heterogeneous information network and unsupervised feature selection.

\subsection{Heterogeneous Information Network}
Meta-path based citation recommendation methods use the citation relationship as supervision to weight different meta-paths \cite{Ren14}. In classification tasks, the importance of different meta-paths can be learned by the informativeness of links \cite{Kong12} \cite{Kong13a}. In recommendation tasks, the implicit user feedback is used as supervision to learn the weights of meta-paths \cite{Yu13} \cite{Yu14}. For link prediction tasks, the existence of links provides guidance to learn importance for different meta-paths \cite{Zhang14a}. When used in information fusion such as cross-network mapping \cite{Kong13} \cite{Zhang14}, the importance of meta-path weights is learned under the supervision of anchor link.

In semi-supervised clustering, PathSelClus \cite{Sun12} requires user guidance to weight different meta-paths for clustering HIN and the HIN can be clustered in different ways based on the user input. SemiRPClus also learns the importance of meta-paths based on labeled information for performing semi-supervised clustering \cite{Luo14}.

However, how to select informative meta-paths in unsupervised scenario has received little attention. Existing work on unsupervised task with meta-paths (e.g., clustering) typically use all the meta-paths generated from the network \cite{Wang15}. The performance of such an approach might be affected by low-quality meta-paths and suffers from poor interpretability.

\subsection{Unsupervised Feature Selection}

In unsupervised feature selection, different heuristics have been explored for selecting features. Selecting features by their spectral property is a popular class of approaches \cite{He05} \cite{Zhao07}. However, these simple heuristics can only evaluate features individually and ignore the correlation among features. Recent methods attempt to overcome this issue by evaluating the subset of features as a whole. Notably, $L_{2, 1}$ norm based methods \cite{Yang11} \cite{Li12} \cite{Qian13} \cite{Du15} have gained much popularity among others. The feature selection problem is performed jointly with linear subspace learning/linear regression. In such methods, the features are evaluated by their utility in the regression problem. Sparsity-inducing $L_{2,1}$ norm is employed to enforce the weights of less useful features shrink to zero. For example, Non-negative Discriminative Feature Selection (NDFS) \cite{Li12} performs non-negative spectral analysis and feature selection jointly. Robust Unsupervised Feature Selection (RUFS) \cite{Qian13} and Robust Spectral Feature Selection (RSFS) \cite{Shi14} study feature selection robust to outlier instances by using $L_{2, 1}$ norm and Huber loss, respectively. FSASL \cite{Du15} employs adaptive structure learning to be more resilient to the noise in the local structure. However, these approaches can only be applied to feature vectors and is not applicable to meta-path selection.

\section{Preliminaries}

In this section, we present some preliminary concepts used in this paper. 

\begin{mydef}
	{\bf Heterogeneous Information Network} The complex side information of data instances can be represented as a {\it Heterogeneous Information Network (HIN)} $\mathcal{G}=(\mathcal{V},\mathcal{E})$. $\mathcal{V}$ denotes the set of nodes, which includes $t$ types of entities, $\mathcal{V}_1 = \{v_{11}, v_{12}, \dots, v_{1n_1}\}$, $\dots$, $\mathcal{V}_t = \{v_{t1}, v_{t2}, \dots, v_{tn_t}\}$. $\mathcal{E}$ denotes the set of (multiple types of) links $\mathcal{E} \subset V\times V$.
\end{mydef}
HIN models the heterogeneous relationship among inter-connected objects. There are various types of real-world data that can be represented as heterogeneous information networks:
\begin{itemize}
	\item Blog network: From a blog user network (Figure \ref{fig:blog}), one could extract the following four types of relationships: {\it user} writes {\it blog post}, which has associated {\it tag} and {\it term}. Besides, {\it user}s are connected with each other by friendship links. Other social media data, such as Twitter and Flickr, can be represented as heterogeneous network in similar manner.
	\item Bibliographic network (Figure \ref{fig:dblp}): there are four types of entities: {\it author}, {\it venue}, {\it paper}, {\it term}, where {\it paper} contains {\it terms}, is written by {\it author} and gets published in certain {\it venue}. Also, a {\it paper} could cite other {\it papers}.
	\item Bioinformatic network: HIN can also represent different entities involved in biological processes. For example, certain disease may be caused by some genes and can be cured by certain chemical compound, which could cause side effects. Such interactions between gene, pathway and chemical compound can be represented as HIN (Figure \ref{fig:gene}).
\end{itemize}

For the type of nodes on which one want to perform machine learning task, we refer to them as {\bf target nodes} in the heterogeneous information network. For example, if the goal is to cluster blog posts, the {\it blog post} nodes are the target nodes in the blog network.

To extract knowledge from HIN, a popular approach is to generate {\it meta-paths} which is defined as follows.

\begin{mydef}
	{\bf Meta-path} A meta-path $\mathcal{P}$ of length $l$ is a sequence of relations $\mathcal{R}_i$ ($i=1, \dots, l$), i.e.,  $\mathcal{T}_1 \xrightarrow{\mathcal{R}_1} \mathcal{T}_2 \xrightarrow{\mathcal{R}_2} \cdots \xrightarrow{\mathcal{R}_l} \mathcal{T}_{l+1}$, where $\mathcal{T}_i$ ($i=1, \dots, l+1$) are the types of nodes. A unique sequence of nodes is referred to as a {\bf path instance} of $\mathcal{P}$.
\end{mydef}

For each pair of nodes, various meta-paths can be extracted to provide information about their correlations from different perspectives. Each meta-path usually carries certain semantics between instances. For example, {\it paper-author-paper} links the papers written by the same author and {\it paper-venue-paper} connects the papers appearing in the same conference. While papers connected by either meta-path are likely to be in the same research area, the former meta-path tends to contain finer grained information. Examples of meta-paths on different HINs can be found in Table \ref{table:metapath}.

A typical way of utilizing the meta-paths is to derive certain similarity/affinity measure from them. Inspired by the path-counting measure in \cite{Sun11}, we define the following side information-based (asymmetric) affinity measure by counting the meta-path instances between the target data points.
\begin{mydef}
	{\bf Max-normalized Meta-path Count} Given a side information network, we define the following affinity measure from the side information w.r.t meta-path $m \in M$ as follows:
	\begin{equation}
	s_{ij}^{(m)} = \frac{|\mathcal{P}^{(m)}(i\leadsto j)|}{\max_{k \neq i} (|\mathcal{P}^{(m)}(i\leadsto k)|)}
	\end{equation}
\end{mydef}
where $|\mathcal{P}^{(m)}(i\leadsto j)|$ denotes the number of path instances with type $m$ between data instances $i$ and $j$, and $|\mathcal{P}^{(m)}(i\leadsto \cdot)|$ denotes the number of out-going path instances of type $m$ from instance $i$. This metric is similar to PathSim \cite{Sun11} in spirit, but PathSim is only applicable for symmetric meta-paths. And we use max-normalization to better preserve the semantic information from our experiment.

Since each meta-path reveals partial information to the correlation between two nodes, combining them together into an aggregated measure provides a more comprehensive view of the correlation. Assuming there are $M$ meta-paths of interest w.r.t. certain type of target node: $\mathbb{P}^{(1)}$, $\mathbb{P}^{(2)}$, $\dots$, $\mathbb{P}^{(M)}$, we can define the following aggregated affinity.

\begin{mydef}
	{\bf Aggregated Meta-path Affinity} For target type $t$ in a heterogeneous information network, we can aggregate the normalized meta-path count of all the meta-paths of interest, into an {\it aggregated meta-path affinity} as follows:
	\begin{equation}
	A_{ij} = \sum_{m=1}^M s_{ij}^{(m)}
	\end{equation}
	where $i, j \in \{1, 2, \dots, n_t\}$.
\end{mydef}
If two nodes are connected by many meta-paths, it indicates they are highly correlated the and the aggregated affinity between them tends to be large.

It should be noted that some meta-paths might be of lower-quality than others. Also, different meta-paths could contain overlapping or redundant information. For instance, one could derive the following meta-paths related to social network with different levels of proximity: {\it Blog-User-User-Blog}, {\it Blog-User-User-User-Blog}, {\it Blog-User-User-User-User-Blog} and so on. We denote them as $BU^2B$, $BU^3B$ and $BU^4B$, respectively. $BU^2B$ captures the first-order proximity between users, which represents the correlation between the blogs written by users who are friends. When the network is sparse, it is desirable to incorporate the second order proximity among users (i.e., friends of friends $BU^3B$). One could extract meta-path with even higher length, such as $BU^4B$ and $BU^5B$. All these meta-paths attempt to exploit the homophily effect of social network and hence carry similar semantic. So, there exists certain redundancy among these meta-paths and it might not be necessary to use all of them. Besides, the utility of these meta-paths is not the same. It is helpful to reduce the number of meta-paths to a succinct subset of meta-paths, which could potentially improve the subsequent machine learning tasks and enhance interpret ability. Therefore, we define the following meta-path reduction problem.

\begin{mydef}\label{def:Reduction}
	{\bf Meta-path Reduction Problem} Our goal is to reduce the $M$ meta-paths set to a $D$ meta-paths subset, where $D < M$. We use $\mathbf{w} \in \{0, 1\}^M$ ($i = 1, \dots, M$) as an indicator vector: $w_m = 1$ indicates the $m$-th meta-path is selected and $w_m=0$ otherwise.
\end{mydef}
As supervision information is not always available, (e.g., in clustering analysis), we aim to propose an effective approach, which generate a reduced subset of $D$ meta-paths that can preserve most of information of all the meta-paths.

\begin{figure*}[t]
	\includegraphics[width=6in]{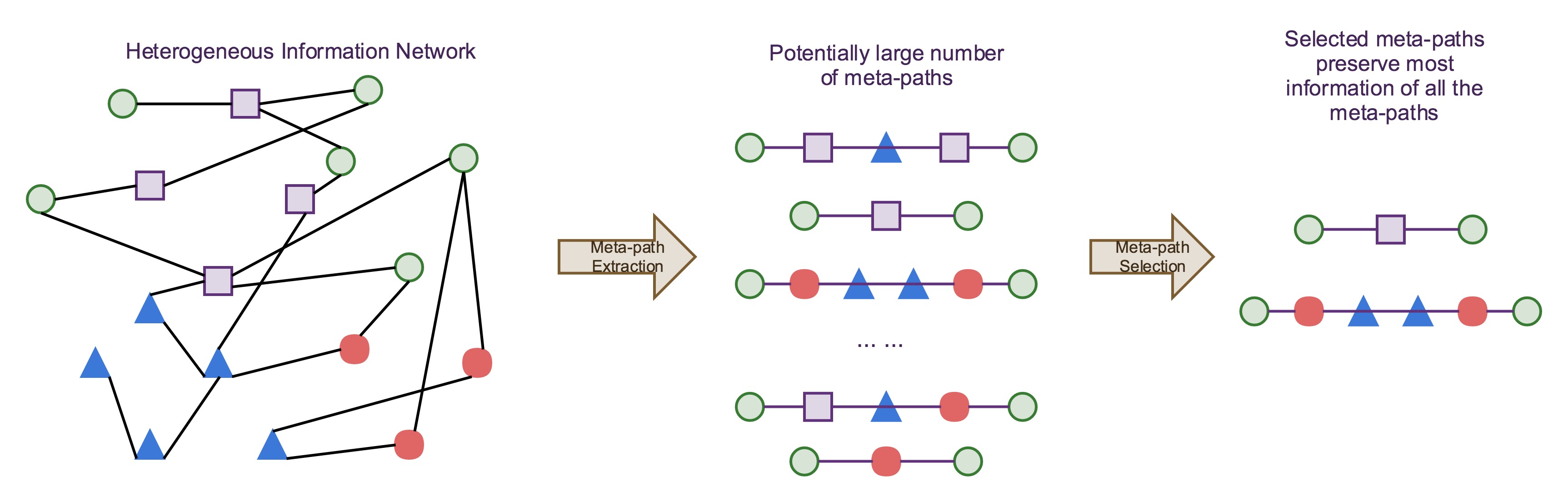}
	\centering
	\caption{Meta-path extraction and meta-path reduction. We select a reduced subset of meta-paths.}
	\label{fig:msframework}
\end{figure*}



\section{Semantic Preserving Meta-path Reduction}

In this section, we present in detail our approach for unsupervised meta-path reduction.

\subsection{Formulations}

Suppose there are $n$ target nodes $v_1, v_2, \dots, v_n$. First, we can utilize the meta-path based affinity to define transition probability between the network nodes. Let us denote the transition probability from $\mathbf{v}_i$ to $\mathbf{v}_j$ ($j \neq i$) as $p_{ij}$ and assume $p_{ij}$ depends on their aggregated affinity $A_{ij}$. Then we can use the softmax function to define this probability.
\begin{mydef}
	{\bf Meta-path based Transition Probability} The following transition probability can be derived from the meta-path based affinity:
	\begin{equation}
	p_{ij} = \frac{\exp{(A_{ij})}}{\sum_{k \neq i} \exp{(A_{ik})}}
	\end{equation}
	where $\sum_{j = 1}^n p_{ij} = 1$.
\end{mydef}
The larger the affinity $A_{ij}$, the larger transition probability $p_{ij}$. We also define self-transition probability $p_{ii} = 0$ ($\forall i = 1, \dots, n$) for convenience.

After meta-path selection, we can still define the transition probability in a similar manner. Let us denote the aggregated affinity on the selected meta-paths as $a_{ij} = \mathbf{w}^T \cdot {s_{ij}}$, where $\mathbf{w}$  is a column vector as we defined in Definition~\ref{def:Reduction} and $s_{ij}$ is a column vector as $(s_{ij}^{(1)},s_{ij}^{(2)},...,s_{ij}^{(M)})^T$. The transition probability from $v_i$ to $v_j$ after meta-path selection is $q_{ij}$:
\begin{equation}
q_{ij} = \frac{\exp(a_{ij})}{\sum_{k \neq i} \exp({a_{ij}})}
\end{equation}
Note that $q_{ij}$ (or $p_{ij}$) is not only determined by $a_{ij}$ (or $A_{ij}$), but also affected by $a_{ik}$ (or $A_{ik}$, $k = 1, \dots, j-1, j +1, \dots, n$) via the normalization term. Therefore, $q_{ij}$ (or $p_{ij}$) is influenced by the relative value of $a_{ij}$ (or $A_{ij}$) compared with other $a_{ik}$ (or $A_{ik}$).

The transition probability captures the structural information among target nodes, which is also the semantics revealed by meta-paths. To preserve the semantics, we try to make two distributions $\mathbf{q}_i = [q_{i1}, \dots, q_{in}]^T$ and $\mathbf{p}_i = [p_{i1}, \dots, p_{in}]^T$ close by minimizing their KL divergence for each $\mathbf{x}_i$.
\begin{equation}
KL(\mathbf{p}_i || \mathbf{q}_i) = \sum_{j \neq i} p_{ij}\log{\frac{p_{ij}}{q_{ij}}}
\end{equation}
The impact of meta-paths should be measured on all pairs of target nodes. So we retain the set of $d$ meta-path which can minimize the sum of KL divergence between $\mathbf{p}_i$ and $\mathbf{q}_i$ on all the data points.
\begin{align}
\begin{split}
\min_{\mathbf{w}} ~ & \sum_{i=1}^n \sum_{j \neq i} p_{ij}\log{\frac{p_{ij}}{q_{ij}}} \\
s.t. ~ & \sum_{t=1}^M w_m = d \\
~ & w_m \in \{0, 1\}, \forall m = 1, \dots, M
\end{split}
\label{eq:obj}
\end{align}

The goal is that, for nodes densely connected by meta-path instances, we still want them to have large transition probability after meta-path reduction so that the semantics can be preserved. For node pairs with low affinity (i.e., it indicates low correlation), it is desirable to keep them loosely connected with reduced subset of meta-paths. So, by minimizing KL-divergence between $\mathbf{p}_i$ and $\mathbf{q}_i$ for $i=1,\dots, n$, we get a reduced subset of meta-paths, which is indicated by the vector $\textbf{w}$, that lets densely connected nodes still easier transition to each other than loosely connected nodes. Thus, the semantic information can be maximally preserved.

\section{Optimization}

\subsection{Relaxation}
The formulation in Eq~(\ref{eq:obj}) is a '0/1' integer programming problem. When the total number of meta-paths is small, one can simply enumerate all combinations with size $D$ and use the combination that leads to smallest objective function, if he/she intends to generate a subset of $D$ meta-paths. However, when the number of meta-paths is large, such brute force approach is time-consuming to optimize. To make the optimization more efficient, we relax the '0/1' constraint on $w_m ~ (\forall m = 1, \dots, M)$ to real values in the range of $[0, 1]$. Also, we use Lagrangian multiplier re-write the summation constraint $\sum_{t=1}^M w_m = D$
\begin{align}
\begin{split}
\min_{\mathbf{w}} ~ & \sum_{i = 1}^n \sum_{j \neq i} p_{ij}\log{\frac{p_{ij}}{q_{ij}}} + \lambda ||\mathbf{w}||_1 \\
s.t. ~ & 0 \leq w_m \leq 1, \forall m = 1, \dots, M
\end{split}
\label{eq:relax}
\end{align}
where $||\cdot||_1$ is the $L_1$ norm and $\lambda$ is the parameter to control the $L_1$ regularization. Note that $|w_m| = w_m$ since $w_t$ ($\forall t = 1, \dots, D$) is always non-negative.

Now we derive the gradient update formula for SPMR. We denote $\exp{(a_{ij})}$ as $E_{ij}$ and the normalization term $\sum_{k \neq i}\exp{(a_{ik})}$ as $Z_i$. So $q_{ij}$ can be denoted as $E_{ij}/Z_i$.

We denote $KL(\mathbf{p}_i||\mathbf{q}_i)=\sum_{j\neq i} p_{ij} \log{\frac{p_{ij}}{q_{ij}}}$ as $\mathcal{L}_i$. The gradient of $\mathcal{L}_i$ w.r.t $s_{ij}$ can be decomposed into two terms,
\begin{align}
\begin{split}
\frac{\partial \mathcal{L}_i}{\partial s_{ij}} & = - \frac{\partial }{\partial s_{ij}}(p_{ij} \log{q_{ij}}) - \sum_{k \neq j} \frac{\partial }{\partial s_{ij}}(p_{ik} \log{q_{ik}}).
\end{split}
\end{align}

Now we derive the gradients on these two parts,
\begin{align}
\begin{split}
\frac{\partial }{\partial s_{ij}}(p_{ij} \log{q_{ij}}) & = p_{ij} / q_{ij} \cdot \frac{\partial q_{ij}}{\partial E_{ij}}\frac{\partial E_{ij}}{\partial s_{ij}} \\
& = p_{ij} / q_{ij} \frac{ Z_i - E_{ij}}{Z_i^2}\cdot \frac{\partial E_{ij}}{\partial s_{ij}}  \\
& = p_{ij} \frac{1}{A_{ij}} \frac{ Z_i  -  E_{ij}}{Z_i}\cdot \frac{\partial E_{ij}}{\partial s_{ij}} \\
& = p_{ij}\frac{1}{E_{ij}} \cdot \frac{\partial A_{ij}}{\partial s_{ij}} - p_{ij} \frac{1}{Z_i} \cdot \frac{\partial E_{ij}}{\partial s_{ij}},
\end{split}
\end{align}

\begin{align}
\begin{split}
\sum_{k \neq j} \frac{\partial }{\partial s_{ij}}(p_{ik} \log{q_{ik}}) & = \sum_{k \neq j} p_{ik} / q_{ik} \cdot \frac{\partial q_{ik}}{\partial E_{ij}}\frac{\partial E_{ij}}{\partial a_{ij}} \\
& = \sum_{k \neq j} -p_{ik}/q_{ik} \frac{ E_{ik}}{Z_i^2} \cdot \frac{\partial E_{ij}}{\partial a_{ij}} \\
& = \sum_{k \neq j} -p_{ik} \frac{1}{Z_i} \cdot \frac{\partial E_{ij}}{\partial a_{ij}}.
\end{split}
\end{align}
By combining them together, it is able to get the following gradient by observing $p_{ij}+\sum_{k \neq j} p_{ik} = 1$,
\begin{align}
\begin{split}
\frac{\partial \mathcal{L}_i}{\partial a_{ij}} = & -p_{ij}\frac{1}{E_{ij}} \cdot \frac{\partial E_{ij}}{\partial a_{ij}} + p_{ij} \frac{1}{Z_i} \cdot \frac{\partial E_{ij}}{\partial a_{ij}}  \\ & +\sum_{k \neq j} p_{ik} \frac{1}{Z_i} \cdot \frac{\partial E_{ij}}{\partial a_{ij}}\\
= & -(p_{ij}\frac{1}{E_{ij}}  -   \frac{1}{Z_i}) \frac{\partial E_{ij}}{\partial a_{ij}} \\
= & -(p_{ij}  -   q_{ij}).
\end{split}
\end{align}
The gradient of loss function $\mathcal{L}$ w.r.t $w_t$ is calculated as follows,
\begin{align}
\begin{split}
\frac{\partial \mathcal{L}}{\partial w_m} & = \sum_{i=1}^n\sum_{j\neq i}\frac{\partial \mathcal{L}_i}{\partial a_{ij}} \frac{\partial a_{ij}}{\partial w_m} + \lambda\frac{\partial |w_m|}{\partial w_m} \\
& = -\sum_{i=1}^n \sum_{j \neq i} (p_{ij} - q_{ij}) s_{ij}^{(m)} + \lambda.
\end{split}
\label{eq:gradient}
\end{align}


Intuitively, when $v_i$ is more likely to connect to $v_j$ than expected (i.e., $p_{ij}<q_{ij}$), the reduced subset of meta-paths with large $s_{ij}^{(m)}$ would be punished to push them away; when the transition probability from $v_i$ to $v_j$ is smaller than desired (i.e., $p_{ij}>q_{ij}$), $w_m$ is updated to pull them closer. Thus, the semantic information can be preserved as originally. If a meta-path has little contribution in preserving the semantics, its weight tends to converge to 0 with $L_1$ regularization, i.e., it will be reduced.


\subsection{Projected Quasi-Newton Method}

To handle the $[0, 1]$ box constraint in the optimization problem, we employ projected Quasi-Newton Method \cite{Bertsekas82}. The reason why we apply Quasi-Newton method is because for a large HIN, the dimension problem is crucial. And in each iteration, it projects $w_m$ ($\forall m = 1, \dots, M$) to the range of $[0, 1]$ after each gradient update with Eq (\ref{eq:gradient})
\begin{equation}
[\text{Proj}_{[0, 1]}(\mathbf{w})]_m = \text{min}(1, \text{max}(0, w_m)).
\end{equation}

Since larger value of $w_m$ indicates higher importance of meta-path, one can retain all the meta-paths with $w_m$ close to $1$ (e.g., $0.9$). Also, larger $\lambda$ would lead to weights of more meta-paths shrink towards zero and less number of meta-paths with $w_m$ close to $1$. Hence, if the goal is to select $D$ meta-paths, he/she could choose the appropriate $\lambda$ that makes $\sum_{m=1}^M I(w_m>0.9) = D$, where $I$ is an indicator function such that when $w_m>0.9$, its value is equal to $1$. We adopt this approach to set $\lambda$ for SPMR in the following experiments.

\section{Experiments}

In this section, we evaluate the proposed SPMR on two real-world datasets.

\subsection{Datasets}

We use the following two datasets:
\begin{itemize}
	\item BlogCatalog~\cite{wang2010discovering}\footnote{\label{blogcatalog}\url{http://dmml.asu.edu/users/xufei/datasets.html}}: A subset of blog post dataset in the following categories: \{Personal Development, Investing, Fitness, Soccer, Cars\}. The heterogeneous network contains users (U), blog posts (B), words (W) and tags (T) as nodes. The dataset contains around 90,000 users with social network. Blog posts are used as target nodes in the experiments.
	\item DBLP: we use the 'four area' dataset in \cite{Sun09} and \cite{Ji10}, which contains {\it author}, {\it paper}, {\it term} and {\it conference} in the following areas: Data Mining, Database, Information Retrieval and Artificial Intelligence. Five representative conferences are selected for each area and a total of $20$ conferences are used. All the papers terms in the paper titles are used to construct the network. The original dataset contains $14376$ papers (P), $14475$ authors (A) and $13571$ terms (T). However, only $4057$ authors have ground-truth labels, we only use these authors as target nodes.
\end{itemize}


\subsection{Experimental Setting}

Similar to unsupervised feature selection \cite{Li12} \cite{Qian13} \cite{Wei16}, we evaluate the quality of selected meta-paths by their clustering performance. Accuracy and Normalized Mutual Information (NMI) are used to evaluate the quality of clustering. Accuracy is defined as follows.
\begin{equation}
Accuracy = \frac{1}{n} \sum_{i=1}^n \mathcal{I}(c_i= map(p_i))
\end{equation}
where $p_i$ is the clustering result of document $i$ and $c_i$ is its real class label. $map(\cdot)$ maps each cluster label to a class label by using Kuhn-Munkres Algorithm \cite{Kuhn55}.

Normalized Mutual Information (NMI) is information theory-based metric for evaluating clustering performance. Let us denote the set of clusters from the ground truth as $C$ and cluster labels obtained from a clustering algorithm as $C'$. Their mutual information $MI(C, C')$ can be defined as follows:
\begin{equation}
MI(C, C') = \sum_{c_i \in C, c'_j \in C'} p(c_i, c'_j) \log \frac{p(c_i, c'_j)}{p(c_i)p(c'_j)}
\end{equation}
where $p(c_i)$ and $p(c'_j)$ are the probabilities that a random instance from the data set belongs to $c_i$ and $c'_j$, respectively, and $p(c_i, c'_j)$ is the joint probability that the instance belongs to the cluster $c_i$ and $c'_j$ simultaneously. In our experiments, we use the normalized mutual information as in previous work \cite{Li12}.
\begin{equation}
NMI(C, C') = \frac{MI(C, C')}{max(H(C), H(C'))}
\end{equation}
where $H(C)$ and $H(C')$ are the entropy of $C$ and $C'$. Higher value of NMI indicates better quality of clustering.

To validate the effectiveness of our proposed methods, we compare two baseline methods, which are using all meta-paths and \textbf{r}andomly \textbf{s}electing (RS) $k$ meta-paths. 



\begin{table}[!htbp]
	\begin{center}
		\caption{Clustering accuracy on two datasets.}
		\begin{tabular}{| l || l | l | l |}
			\hline
			Dataset & \multicolumn{3}{c|}{{\bf BlogCatalog}} \\ \hline
			\# meta-paths   & 3 & 6 & 9   \\ \hline
			All paths (15) & \multicolumn{3}{c|}{0.3975}   \\ \hline
			RS  & 0.4171 & 0.4191 & 0.4454   \\ \hline
			SPMR & {\bf 0.5661} & 0.4472 & 0.5155  \\ \hline
			\hline
			Method & \multicolumn{3}{c|}{{\bf DBLP}} \\ \hline
			\# meta-paths   & 1 & 3 & 5   \\ \hline
			All paths (6) & \multicolumn{3}{c|}{0.3485} \\ \hline
			RS  & 0.3337 & 0.3313 & 0.3356   \\ \hline
			SPMR & 0.3495 & {\bf 0.3544} & 0.3341    \\ \hline
			\hline
		\end{tabular}
		\label{table:acc}
	\end{center}
\end{table}

\begin{table}[!htbp]
	\begin{center}
		\caption{Clustering NMI on two datasets.}
		\begin{tabular}{| l || l | l | l |}
			\hline
			Dataset & \multicolumn{3}{c|}{{\bf BlogCatalog}} \\ \hline
			\# meta-paths   & 3 & 6 & 9   \\ \hline
			All paths (15) & \multicolumn{3}{c|}{0.1861}   \\ \hline
			RS  & 0.2126 & 0.2154 & 0.2550   \\ \hline
			SPMR & {\bf 0.3777} & 0.2716 & 0.3225  \\ \hline
			\hline
			Method & \multicolumn{3}{c|}{{\bf DBLP}} \\ \hline
			\# meta-paths   & 1 & 3 & 5   \\ \hline
			All paths (6) & \multicolumn{3}{c|}{0.0494} \\ \hline
			RS   & 0.0393 & 0.0326 & 0.0381   \\ \hline
			SPMR & 0.0536 & {\bf 0.0578} & 0.0402    \\ \hline
			\hline
		\end{tabular}
		\label{table:nmi}
	\end{center}
\end{table}

\subsection{Results}

The clustering results on two datasets with different numbers of selected meta-paths are shown in Table \ref{table:acc} and Table \ref{table:nmi}, w.r.t. accuracy and NMI respectively. We can observe that the clustering performance can usually be improved with a reduced set of meta-paths, compared with using all meta-paths and random selecting meta-paths. 

For example, on BlogCatalog dataset, for the accuracy of clustering, using $3$ selected meta-paths outperforms using all paths by $42.4\%$ and randomly selecting $3$ meta-paths by $35.7\%$. On DBLP dataset, with respect to NMI, using $3$ meta-path improves $17.0\%$ compared with using all the meta-paths and $77.3\%$ compared with randomly selecting $3$ meta-paths. This indicates the usefulness of performing meta-path selection for unsupervised task.


\begin{table}
	\begin{center}
		\caption{Reduced Subset of Meta-paths on BlogCatalog}
		\resizebox{0.35\textwidth}{!}{
			\begin{tabular}{| l | l | }
				\hline
				\# & Ranked meta-paths \\ \hline
				\multirow{3}{*}{3} &
				Blog - User - Blog - Tag - Blog \\
				& Blog - Tag -  Blog - Tag - Blog \\
				& Blog - User - User - Blog - Tag - Blog  \\ \hline
				\multirow{6}{*}{6} & Blog - User - Blog - User - User - Blog \\
				& Blog - User - Blog - Tag - Blog \\
				& Blog - User - Blog - User - Blog \\
				& Blog - User - User - Blog - Tag - Blog \\
				& Blog - Tag - Blog - Tag - Blog \\
				& Blog - Tag - Blog - Words - Blog \\ \hline
				\multirow{9}{*}{9} & Blog - User - Blog - User - User - Blog \\
				& Blog - User - Blog - Tag - Blog \\
				& Blog - User - Blog - Words - Blog \\
				& Blog - User - Blog - User - Blog \\
				& Blog - User - User - Blog - Tag - Blog \\
				& Blog - User - User - User - Blog \\
				& Blog - Tag - Blog - Tag - Blog \\
				& Blog - User - User - Blog - Tag - Blog \\
				& Blog - Tag - Blog \\  \hline
			\end{tabular}
		}
		\label{table:blog_mp}
	\end{center}
\end{table}


Table \ref{table:blog_mp} lists the selected meta-paths on BlogCatalog dataset. Selecting three meta-paths on BlogCatalog achieves the best performance, and the reduced set of meta-paths does not include word-related meta-paths. This suggests that meta-paths derived from user and tag tend to best preserve the semantic information. Additionally, from the reduction process of meta-paths, we can know the characteristics of BlogCatalog heterogeneous network. For example, the elimination of {\it Blog - User - User - User - Blog} meta-path suggests that the friends of friends relationship might not provide useful information of BlogCatalog, which can reveal the sparsity of BlogCatalog.

Table \ref{table:dblp_mp} lists the selected meta-paths on dblp dataset, the performance of a single meta-path {\it Author - Paper - Author - Paper - Term - Paper - Author} is similar to that of all the meta-paths, which reveals that this longest meta-path can preserve most of the semantic information. Also, the {\it Author - Paper -Term - Paper - Term - Paper - Author } is dropped as the result of redundancy. Other mate-paths can also contain this kind of semantics because the second paper must have a author, then this {\it Author - Paper - Term - Paper - Term - Paper - Author} meta-path can also be detected by {\it Author - Paper - Term - Paper - Author} meta-path.


 \begin{table}
 \begin{center}
     \caption{Selected meta-paths on DBLP}
     \resizebox{0.49\textwidth}{!}{
     \begin{tabular}{| l | l | }
     \hline
     \# & Ranked meta-paths \\ \hline
     \multirow{1}{*}{1} & {Author - Paper - Author - Paper - Term - Paper - Author}\\ \hline
     \multirow{3}{*}{3} & Author - Paper - Author - Paper - Author \\
                 & Author - Paper - Author - Paper - Term - Paper - Author \\
                 & Author - Paper - Term - Paper - Author  \\ \hline
     \multirow{5}{*}{5} & Author - Paper - Author - Paper - Author \\
                 & Author - Paper - Author - Paper - Author - Paper - Author \\
                 & Author - Paper - Author - Paper - Term - Paper - Author \\
                 & Author - Paper - Term - Paper - Author \\
                 & Author - Paper - Term - Paper - Term - Paper - Author \\ \hline
     \end{tabular}
     }
     \label{table:dblp_mp}
 \end{center}
 \end{table}




\section{Conclusion}

From heterogeneous information networks, one could extract many meta-paths, but some meta-paths contains misleading noise or redundant information. Hence, applying reduction on the number of meta-paths, the performance of subsequent data mining tasks could be improved. Also the reduced subset of meta-paths can reveal the hidden characteristic of HIN. As supervision information is not always available, we study the problem of meta-path reduction in unsupervised setting. We propose a new method which aims to preserve the transition probability so that the semantics can be preserved. An optimization method based on projected Quasi-Newton method is proposed to solve the optimization problem. Experimental results shows the proposed SPMR can reduce the number of meta-paths in unsupervised setting, while preserving the semantic information, hence can enhance the performance of unsupervised task.

\clearpage

\bibliography{reference}

\begin{thebibliography}{}

\bibitem[\protect\citeauthoryear{Bertsekas}{1982}]{Bertsekas82}
Bertsekas, D.~P.
\newblock 1982.
\newblock Projected newton methods for optimization problems with simple
  constraints.
\newblock In {\em SIAM Jounal on Control and Optimization}.

\bibitem[\protect\citeauthoryear{Du and Shen}{2015}]{Du15}
Du, L., and Shen, Y.-D.
\newblock 2015.
\newblock Unsupervised feature selection with adaptive structure learning.
\newblock In {\em Proceedings of the 21st ACM SIGKDD International Conference
  on Knowledge Discovery and Data Mining},  209--218.

\bibitem[\protect\citeauthoryear{He, Cai, and Niyogi}{2005}]{He05}
He, X.; Cai, D.; and Niyogi, P.
\newblock 2005.
\newblock Laplacian score for feature selection.
\newblock In {\em NIPS}.

\bibitem[\protect\citeauthoryear{Ji \bgroup et al\mbox.\egroup }{2010}]{Ji10}
Ji, M.; Sun, Y.; Danilevsky, M.; Han, J.; and Gao, J.
\newblock 2010.
\newblock Graph regularized transductive classification on heterogeneous
  information networks.
\newblock In {\em ECML/PKDD (1)}, volume 6321,  570--586.

\bibitem[\protect\citeauthoryear{Kong \bgroup et al\mbox.\egroup
  }{2012}]{Kong12}
Kong, X.; Yu, P.~S.; Ding, Y.; and Wild, D.~J.
\newblock 2012.
\newblock Meta path-based collective classification in heterogeneous
  information networks.
\newblock In {\em CIKM},  1567--1571.

\bibitem[\protect\citeauthoryear{Kong, Cao, and Yu}{2013}]{Kong13a}
Kong, X.; Cao, B.; and Yu, P.~S.
\newblock 2013.
\newblock Multi-label classification by mining label and instance correlations
  from heterogeneous information networks.
\newblock In {\em Proceedings of the 19th ACM SIGKDD International Conference
  on Knowledge Discovery and Data Mining},  614--622.
\newblock ACM.

\bibitem[\protect\citeauthoryear{Kong, Zhang, and Yu}{2013}]{Kong13}
Kong, X.; Zhang, J.; and Yu, P.~S.
\newblock 2013.
\newblock Inferring anchor links across multiple heterogeneous social networks.
\newblock In {\em CIKM},  179--188.

\bibitem[\protect\citeauthoryear{Kuhn}{1955}]{Kuhn55}
Kuhn, H.~W.
\newblock 1955.
\newblock {The Hungarian Method for the Assignment Problem}.
\newblock {\em Naval Research Logistics Quarterly} 2(1--2):83--97.

\bibitem[\protect\citeauthoryear{Leskovec, Lang, and
  Mahoney}{2010}]{Leskovec10}
Leskovec, J.; Lang, K.~J.; and Mahoney, M.~W.
\newblock 2010.
\newblock Empirical comparison of algorithms for network community detection.
\newblock In {\em WWW},  631--640.

\bibitem[\protect\citeauthoryear{Li \bgroup et al\mbox.\egroup }{2012}]{Li12}
Li, Z.; Yang, Y.; Liu, J.; Zhou, X.; and Lu, H.
\newblock 2012.
\newblock Unsupervised feature selection using nonnegative spectral analysis.
\newblock In {\em AAAI}.

\bibitem[\protect\citeauthoryear{Liu \bgroup et al\mbox.\egroup
  }{2017}]{liu2017event}
Liu, Z.; Yang, Y.; Huang, Z.; Shen, F.; Zhang, D.; and Shen, H.~T.
\newblock 2017.
\newblock Event early embedding: Predicting event volume dynamics at early
  stage.
\newblock In {\em Proceedings of the 40th International ACM SIGIR Conference on
  Research and Development in Information Retrieval},  997--1000.
\newblock ACM.

\bibitem[\protect\citeauthoryear{Lu and Getoor}{2003}]{Lu03}
Lu, Q., and Getoor, L.
\newblock 2003.
\newblock Link-based classification.
\newblock In {\em ICML},  496--503.

\bibitem[\protect\citeauthoryear{Luo, Pang, and Wang}{2014}]{Luo14}
Luo, C.; Pang, W.; and Wang, Z.
\newblock 2014.
\newblock Semi-supervised clustering on heterogeneous information networks.
\newblock In {\em PAKDD (2)}, volume 8444,  548--559.

\bibitem[\protect\citeauthoryear{Qian and Zhai}{2013}]{Qian13}
Qian, M., and Zhai, C.
\newblock 2013.
\newblock Robust unsupervised feature selection.
\newblock In {\em IJCAI}.

\bibitem[\protect\citeauthoryear{Ren \bgroup et al\mbox.\egroup }{2014}]{Ren14}
Ren, X.; Liu, J.; Yu, X.; Khandelwal, U.; Gu, Q.; Wang, L.; and Han, J.
\newblock 2014.
\newblock Cluscite: effective citation recommendation by information
  network-based clustering.
\newblock In {\em Proceedings of the 20th ACM SIGKDD International Conference
  on Knowledge Discovery and Data Mining},  821--830.

\bibitem[\protect\citeauthoryear{Shi, Du, and Shen}{2014}]{Shi14}
Shi, L.; Du, L.; and Shen, Y.-D.
\newblock 2014.
\newblock Robust spectral learning for unsupervised feature selection.
\newblock In {\em ICDM}.

\bibitem[\protect\citeauthoryear{Sun \bgroup et al\mbox.\egroup }{2011}]{Sun11}
Sun, Y.; Han, J.; Yan, X.; Yu, P.~S.; and Wu, T.
\newblock 2011.
\newblock Pathsim: Meta path-based top-k similarity search in heterogeneous
  information networks.
\newblock {\em PVLDB} 4(11):992--1003.

\bibitem[\protect\citeauthoryear{Sun \bgroup et al\mbox.\egroup }{2012}]{Sun12}
Sun, Y.; Norick, B.; Han, J.; Yan, X.; Yu, P.~S.; and Yu, X.
\newblock 2012.
\newblock Integrating meta-path selection with user-guided object clustering in
  heterogeneous information networks.
\newblock In {\em Proceedings of the 18th ACM SIGKDD International Conference
  on Knowledge Discovery and Data Mining},  1348--1356.

\bibitem[\protect\citeauthoryear{Sun \bgroup et al\mbox.\egroup
  }{2018}]{sun2018mega}
Sun, L.; He, L.; Huang, Z.; Cao, B.; Xia, C.; Wei, X.; and Yu, P.~S.
\newblock 2018.
\newblock Joint embedding of meta-path and meta-graph for heterogeneous
  information networks.
\newblock In {\em IEEE International Conference on Big Knowledge}.
\newblock IEEE.

\bibitem[\protect\citeauthoryear{Sun, Yu, and Han}{2009}]{Sun09}
Sun, Y.; Yu, Y.; and Han, J.
\newblock 2009.
\newblock Ranking-based clustering of heterogeneous information networks with
  star network schema.
\newblock In {\em KDD},  797--806.

\bibitem[\protect\citeauthoryear{Wang \bgroup et al\mbox.\egroup
  }{2010}]{wang2010discovering}
Wang, X.; Tang, L.; Gao, H.; and Liu, H.
\newblock 2010.
\newblock Discovering overlapping groups in social media.
\newblock In {\em Data Mining (ICDM), 2010 IEEE 10th International Conference
  on},  569--578.
\newblock IEEE.

\bibitem[\protect\citeauthoryear{Wang \bgroup et al\mbox.\egroup
  }{2015}]{Wang15}
Wang, C.; Song, Y.; Li, H.; Zhang, M.; and Han, J.
\newblock 2015.
\newblock Knowsim: A document similarity measure on structured heterogeneous
  information networks.
\newblock In {\em ICDM},  1015--1020.

\bibitem[\protect\citeauthoryear{Wei and Yu}{2016}]{Wei16}
Wei, X., and Yu, P.~S.
\newblock 2016.
\newblock Unsupervised feature selection by preserving stochastic neighbors.
\newblock In {\em AISTATS}.

\bibitem[\protect\citeauthoryear{Yang \bgroup et al\mbox.\egroup
  }{2011}]{Yang11}
Yang, Y.; Shen, H.~T.; Ma, Z.; Huang, Z.; and Zhou, X.
\newblock 2011.
\newblock l2, 1-norm regularized discriminative feature selection for
  unsupervised learning.
\newblock In {\em IJCAI},  1589--1594.

\bibitem[\protect\citeauthoryear{Yu \bgroup et al\mbox.\egroup }{2013}]{Yu13}
Yu, X.; Ren, X.; Sun, Y.; Sturt, B.; Khandelwal, U.; Gu, Q.; Norick, B.; and
  Han, J.
\newblock 2013.
\newblock Recommendation in heterogeneous information networks with implicit
  user feedback.
\newblock In {\em RecSys},  347--350.

\bibitem[\protect\citeauthoryear{Yu \bgroup et al\mbox.\egroup }{2014}]{Yu14}
Yu, X.; Ren, X.; Sun, Y.; Gu, Q.; Sturt, B.; Khandelwal, U.; Norick, B.; and
  Han, J.
\newblock 2014.
\newblock Personalized entity recommendation: a heterogeneous information
  network approach.
\newblock In {\em WSDM},  283--292.

\bibitem[\protect\citeauthoryear{Zhang, Kong, and Yu}{2014}]{Zhang14a}
Zhang, J.; Kong, X.; and Yu, P.~S.
\newblock 2014.
\newblock Transferring heterogeneous links across location-based social
  networks.
\newblock In {\em WSDM},  303--312.

\bibitem[\protect\citeauthoryear{Zhang, Yu, and Zhou}{2014}]{Zhang14}
Zhang, J.; Yu, P.~S.; and Zhou, Z.-H.
\newblock 2014.
\newblock Meta-path based multi-network collective link prediction.
\newblock In {\em Proceedings of the 20th ACM SIGKDD International Conference
  on Knowledge Discovery and Data Mining},  1286--1295.

\bibitem[\protect\citeauthoryear{Zhao and Liu}{2007}]{Zhao07}
Zhao, Z., and Liu, H.
\newblock 2007.
\newblock Spectral feature selection for supervised and unsupervised learning.
\newblock In {\em ICML}, volume 227,  1151--1157.

\end{thebibliography}
\bibliographystyle{aaai}

\end{document}